\documentclass[a4paper,12pt]{article}
\usepackage[]{psfig}
\newcommand{\be}{\begin{equation}}
\newcommand{\ee}{\end{equation}}
\newcommand{\ba}{\begin{eqnarray}}
\newcommand{\ea}{\end{eqnarray}}

\begin{document}
\baselineskip 0.65cm
\title{\bf Particle-vortex dynamics in noncommutative space}
\author{D.H.~Correa$^a$\thanks{CONICET} \,,
G.S.~Lozano$^b$\thanks{Associated with CONICET} \, , \\
E.F.~Moreno$^a\dagger$ and
F.A.Schaposnik$^a$\thanks{Associated with CICPBA}\\
{\normalsize\it $^a$Departamento de F\'\i sica, Universidad Nacional
de La Plata}\\
{\normalsize\it C.C. 67, 1900 La Plata, Argentina}
\\
{\normalsize\it $^b$Departamento de F\'\i sica, FCEyN, Universidad de
Buenos Aires}\\
{\normalsize\it Pab.1, Ciudad Universitaria,
CP 1428, Buenos Aires,Argentina}
}

\date{\today}
\maketitle

\begin{abstract}
We  study the problem of a charged particle in the presence of a
uniform magnetic field plus a vortex  in noncommutative planar
space considering the two possible  non-commutative extensions of
the corresponding Hamiltonian,  namely the ``fundamental'' and the
``antifundamental'' representations. Using a Fock space formalism
we construct  eigenfunctions and eigenvalues  finding in each case
half of the states existing in the ordinary space case. In the
limit of $\theta \to 0$ we recover the two classes of states found
in ordinary space, relevant for the study of anyon physics.
\end{abstract}

The Landau and the Aharonov-Bohm problems are the two paradigms
of planar quantum mechanics of charged particles in a magnetic
field \cite{jac}-\cite{jac2}. Inspired by recent observations concerning
the relevance of noncommutative geometry for
describing the Quantum Hall effect \cite{Sus}-\cite{R}, both problems
have been recently considered in noncommutative space.

Concerning the Landau problem -charged particles in a
constant magnetic field-  it  was recently
analyzed  in noncommutative space in
\cite{np}-\cite{BNS}. Related discussions   within the framework
of the extended Galilean group were presented in \cite{luk}-\cite{dh3}.
Regarding the Aharonov-Bohm problem  -charged particles
in the background of a vortex-
some results in noncommutative space were presented in \cite{Cha1}-\cite{G3}.

Adding both magnetic fields, one uniform, the other vortex-like,
turns the problem  even more interesting. To our knowledge this
system was first analyzed and solved in ordinary space in
\cite{JC}, in the context of anyon physics, where particles with
exotic statistic are represented by charges pierced by a magnetic
flux. In this regard, the problem of two anyons in an external
magnetic field is equivalent to that of a particle in the presence
of an external magnetic field plus a vortex. From the approach of
\cite{JC} one concludes that in ordinary space, the eigenstates
and spectrum   can be found analytically  by modifying in a simple
but subtle way, the usual Landau problem ladder operator
formalism. The presence of the vortex manifests itself in a  way
simpler than that of the usual Aharonov-Bohm problem.

We shall consider here the generalization of this problem to
noncommutative planar space. Namely, we shall solve the
Schr\"odinger equation for a charged particle in the presence of a
uniform magnetic field plus a vortex, when space coordinates
satisfy
\be
[x^1,x^2] = i\theta  \,\,\,.
\label{1} \ee
It is convenient to introduce
complex variables $z$ and $\bar z$
\be z = \frac{1}{\sqrt{2}}(x^1 + i x^2)\, , \;\;\;\;\;\; \bar z=
\frac{1}{\sqrt{2}}(x^1 - i x^2) \label{2} \ee
and, associated to them, annihilation and creation operators  $ a$
and $ a^\dagger$ in the form
\be a = \frac{1}{\sqrt{2\theta}}(x^1 + i x^2)\, , \;\;\;\;\;\;
a^\dagger = \frac{1}{\sqrt{2\theta}}(x^1 - i x^2) \,\,\,,
\label{3} \ee
so that (\ref{1}) becomes
\be
[ a, a^\dagger] = 1 \,\,\, .
\label{4}
\ee
In this way, through the action of $a^\dagger$ on the vacuum state
$|0\rangle$, eigenstates of the number operator
\be
 N = { a}^\dagger  a
\label{number} \ee
are generated.With this conventions, derivatives in the Fock space
are given by
\be
\partial_z = -\frac{1}{\sqrt \theta}
[ a^\dagger,~] \, , \;\;\;\;\;\; \partial_{\bar z} =
\frac{1}{\sqrt \theta} [ a,~] \label{5}  \,\,\,.   \ee
Some useful formul{\ae} for what follows are
\ba
a^q f(N) = f(q+N) a^q \nonumber \; , &&
\left(a^\dagger\right)^q f(N) =f(N-q) \left(a^\dagger\right)^q \nonumber \\
 \left[a^\dagger,a ^p \right]  = - pa^{(p-1)}
 \; , & &
 \left[a,
 (a^\dagger)^p\right]  =  p
 \left(a^\dagger \right)^{p-1}
\,\,\,.
\ea

As stated above, we shall discuss the quantum mechanics of a particle
in the background of a constant magnetic field plus a magnetic vortex.
For the sake of clarity, we first analyze the constant field strength alone
and then discuss the effect of adding a vortex field.
We define the vector potential components
\be
 A_z = \frac{1}{\sqrt {2}} ( A_1 - i  A_2) \, ,
\;\;\;\;  A_{\bar z} = \frac{1}{\sqrt {2}} ( A_1 + i
A_2) \,\,\, ,
\label{vec}
\ee
so that the field strength can be written as
\be
{F}_{z\bar z} =
\partial_z A_{\bar z} - \partial_{\bar z} A_z +ie[A_z,A_{\bar z}]
\label{F}
\,\,\, .
\ee
A constant  field strength background is given by
\be
{ F}_{z\bar z} = iB^0 \,\,\, .
\label{bac}
\ee
Let us now introduce covariant derivatives. In noncommutative space,
even for an Abelian gauge theory
one has the possibility of defining them in ``fundamental'' and
``anti-fundamental'' representations,
\begin{eqnarray}
 D_z \Psi= \partial_z \Psi+ i e A_z\Psi & & D_{\bar z}
 \Psi= \partial_{\bar z} \Psi+
 i e A_{\bar z}\Psi
\;\;\;\; {\rm fundamental} \nonumber
 \\
 D_z \Psi= \partial_z \Psi- i e \Psi A_z & & D_{\bar z}
 \Psi= \partial_{\bar z}\Psi -
 i e \Psi
 A_{\bar z}
\;\;\;\; {\rm anti\!-\!fundamental}
 \,\,\, .\label{deris}
\end{eqnarray}
There is also the possibility of defining the covariant derivative
in the adjoint representation, which will not be considered here.
We shall first discuss the case of the fundamental representation
and then extend the results to the anti-fundamental.

The Schr\"odinger equation governing the time evolution of the wave
function $\Psi$ is given by
\be
-\frac{1}{2m}
\left[D_z ,D_{\bar z} \right]_+ \Psi =
i\frac{\partial \Psi}{\partial t}
\label{scho}
\ee
where we have defined $[f,g]_+ = fg + gf$. Then writing
\be
\Psi = \Psi(\bar z,z) \exp (-iEt) \,\,\, ,
\label{est}
\ee
one has, for the time-independent  Schr\"odinger equation,
\be
H \Psi(\bar z, z)
=-\frac{1}{2m}
\left[D_z ,D_{\bar z} \right]_+ \Psi
 =
E \Psi(\bar z,z)
\,\,\, .
\label{scho2}
\ee
Let us now construct the angular momentum operator $L$. To this
end, it is useful to define the ``Eulerian''  position operators
$w$ and $\bar{w}$, related to the ``Lagrangian'' ones ($z$ and
$\bar z$) through \cite{Sus}
\be
w=z+ie\theta A_{\bar z} \;\;\;\; \bar{w}=\bar{z}-ie\theta A_z
\,\,\, .
\ee
Given the angular momentum operator, defined in ordinary space as
\be
L = z \partial_z - \bar z \partial_{\bar z}  \; ,
\label{dieci}
\ee
one can see that in noncommutative space  the
appropriate $L$-operator
in the
presence of a constant magnetic field $B^0$
can be written  as \cite{np},\cite{cor2}
\be
L =  wD_z -\bar{w} D_{\bar z}
   - \frac{eB^0}{2\left(1 - eB^0\theta\right)}
\left[\bar w, w  \right ]_+
- \frac{\theta}{2} \left[ D_{\bar z},D_z \right]_+
\label{angu}
\,\,\, .
\ee
It is important to note that  $L$, as defined in (\ref{angu}) is
the generalization to noncommutative space of the so called
``mechanical angular momentum''  \cite{J}.

It is convenient at this point to introduce two pairs of creation
and annihilation operators, the noncommutative analogue of those
developed for the usual Landau problem \cite{Mac},\cite{JC} (for
the constant magnetic field case, a related discussion in
noncommutative space is presented in
\cite{HR},\cite{np}-\cite{BNS}). The actual form of these
operators depend  on the sign of $eB^0$ and, in the present
noncommutative case,  also on the magnitude of $eB_0 \theta$. They
are defined as
\be
c_1 =
\frac{1}{\sqrt{|eB^0|}}
D_z
\; , \;\;\; \; \;\;\; \;\;\; \; \;\;\; \;\;\; \; \;\;\;
c_2 =
-\frac{1}{\sqrt{|eB^0|}}
D_{\bar z}
\ee
\be
d_1 = \frac{1}{\sqrt{|\kappa|}}  w +
 {\rm sgn}\kappa \sqrt{|\kappa|} D_{\bar z}
 \; , \;\;\; \; \;\;\;
 d_2 = \frac{1}{\sqrt{|\kappa|}} \bar w -
 {\rm sgn}\kappa \sqrt{|\kappa|} D_z
\,\,\, ,
\label{alpha}
\ee
where
\be
\kappa = \frac{1 - eB^0\theta}{eB^0}
\,\,\, .
\label{kappa}
\ee
One can verify that with this definition
\be
[c_1,c_2] = {\rm sgn} (eB^0) \; , \;\;\;   [d_1,d_2] =
{\rm sgn}\, \kappa \; ,
\;\;\;
 [c_i, d_j] =  0
 \label{algebra1}
\ee
Since
\be
c_2 = c_1^\dagger \; , \;\;\;\; d_2 = d_1^\dagger
\,\,\, ,
\label{muta}
\ee
operators $c_1$ and $c_2$  will play, depending on the sign of
$eB^0$ and $\kappa$, the role of creation ($c^\dagger$)  or
annihilation ($c$) operators and the same occurs for the $d_i$'s.

In terms of these operators, the Hamiltonian  (\ref{scho2}) takes
the form
\be
H = \frac{\omega}{2} \left[c, c^\dagger \right]_+
\label{hami}
\ee
with
\be
 \omega = \frac{|eB^0|}{m}
\,\,\, .
 \label{omega}
 \ee
Concerning the angular momentum, one has to distinguish two regimes,
\ba
L &=& {\rm sgn}\, \kappa \,( c^\dagger c - d^\dagger d)   \;\;\;\; {\rm for}
~ ~ ~ ~ 1 - eB^0\theta > 0 \,\,\, , \nonumber\\
L &=&  -  \,( c^\dagger c + d^\dagger d)   ~ ~ ~ ~ \;\;\;\; {\rm for}
~ ~ ~ ~ \,
1 - eB^0\theta < 0 \,\,\, .\label{angust}
\ea
In both cases, one can easily see that
\be
[H,L] = 0
\,\,\, .
\ee

In order to determine the spectrum one has then to find the common
eigenfunctions of $H$ and $L$, that is, one has to construct the Fock
space of $c$'s and $d$'s. We shall now proceed to this construction but
for the more general case in which a magnetic vortex is added to the
constant magnetic field background. That is, we shall consider a
field strength of the form
\be
 F_{z \bar z}
 = i B^0 -
 i\frac{\alpha}{e\theta} |0\rangle\langle 0 |
\,\,\, .
\label{bac2}
 \ee
Here $|0\rangle\langle 0 |$ is the projector onto the  state
$|0\rangle$ annihilated by operator $a$ defined in eq.(\ref{3}).
One can easily see that in the $\theta \to 0$ limit, the second
term in (\ref{bac2}) goes, in configuration space, to a delta
function corresponding to a singular vortex at the origin, with
flux related to the real parameter $\alpha$ according to
\cite{POL}-\cite{JKL}
\be
\Phi^{vor} = 2\pi i\theta \,{\rm Tr} F^{vor}_{z\bar z} = \frac{2 \pi}{e} \alpha
\,\,\, .
\label{flujo}
\ee

As in the constant field case, we can  introduce operators $c_i$ and
$d_i$ now taking the form

\be
c_1 =
\frac{1}{\sqrt{|eB^0|}}
D_z[A]
\; , \;\;\; \; \;\;\; \;\;\; \; \;\;\; \;\;\; \; \;\;\;
c_2 =
-\frac{1}{\sqrt{|eB^0|}}
D_{\bar z}[A]
\label{alpha1}
\ee
\be
d_1 = \frac{1}{\sqrt{|\kappa|}}  w[A] +
 {\rm sgn}\kappa \sqrt{|\kappa|} D_{\bar z}[A]
 \; , \;\;\; \; \;\;\;
 d_2 = \frac{1}{\sqrt{|\kappa|}} \bar w[A] -
 {\rm sgn}\kappa \sqrt{|\kappa|} D_z[A]  \,\,\, ,
\label{alpha2}
\ee
where $\kappa$ is still given by (\ref{kappa}) (i.e.,  depending
only on the constant part of the magnetic field). Concerning the
operator algebra, one has, instead of (\ref{algebra1}),
\ba
[c_1,c_2] &=& {\rm sgn} (eB^0)  - \frac{\alpha}{\theta|eB^0|}
|0\rangle\langle 0| \,\,\, ,\nonumber\\
      {[d_1,d_2]} &=&  {\rm sgn}\, \kappa
 +  \frac{\alpha}{\theta|eB^0(1-eB^0\theta)|}
  |0\rangle\langle 0| \,\,\, , \nonumber\\
  {[c_1, d_1] } &=& \frac{\alpha}{eB^0\theta\sqrt{|1-eB^0\theta|}}
  |0\rangle\langle 0|  \,\,\,  , \\
{[c_2, d_2]}  &=& -\frac{\alpha}{eB^0\theta\sqrt{|1-eB^0\theta|}}
|0\rangle\langle 0| \,\,\, ,\nonumber\\
\label{algebra2} {[c_1,d_2]} &= &{[c_2,d_1]} = 0 \,\,\, .
\ea
Note that for those states $\chi$  such that
\be
 | 0 \rangle \langle 0 |\chi= 0 \; , \;\;  \;\;\;
\label{formita}
\ee
this algebra coincides with that defined in (\ref{algebra1}). We
shall call ${\cal P}_0$ the subspace of states satisfying
(\ref{formita}).  This condition, in the commutative space limit
becomes the ``hard-core condition'' since it corresponds to the
vanishing of wave functions at the origin. Moreover, in the
$\theta \to 0$ limit , the algebra (\ref{algebra2}) coincides with
that obtained in \cite{JC} (except that in this last reference it
is presented in the singular gauge).

Again, depending on the signs and magnitudes of $eB^0$ and $\theta$
the $c_i$'s and $d_i$'s will act as creation or annihilation
operators. As in the constant magnetic field case, the Hamiltonian $H$
and angular momentum $L$ can be written in the form
(\ref{hami}) and (\ref{angu}), \underline{provided states are restricted
to ${\cal P}_0$}.

At this point we have to write an explicit expression for the
vector potential leading to a  field strength $F_{z\bar z}$ as
given by (\ref{bac2}). One can see that a possible choice is
\be
A_z=\frac{i}{\sqrt \theta} g( N)  a^\dagger \;\; ,\;\; A_{\bar z}=-
\frac{i}{\sqrt \theta}
 a g( N)
\ee
with
\be
g( N)=-\frac{1}{ e}
\left(1-\sqrt{1 - {eB^0}{\theta} +\frac{\alpha}{ N}}\right)
\,\,\, .
\ee
Here, we have proposed a form which is valid for $1 - {eB^0}{\theta} >0$
and positive $\alpha$. Other possibilities can be handled similarly.

Given the Hamiltonian (\ref{hami}) and angular momentum
(\ref{angust}),
we shall now construct the Fock space associated to the operators
$N_c =   c^\dagger c $ and $N_d = d^\dagger d$. Let us start by
considering a  state $\chi$ such that
\be
c \chi = 0 \,\, \,.
\label{could}
\ee
Then $\chi$ is an eigenstate of the Hamiltonian
\be
H \chi = \frac{\omega}{2} \chi
\,\,\, .
\label{chis}
\ee
In order to make $\chi$ also an eigenstate of $L$, we propose the following
ansatz
\be
\chi(a,a^\dagger)= { a}^{\dagger n } h( N)
\,\,\, .
\label{ansatz+}
\ee
Note that in the  commutative  limit $a^{\dagger n}$ can be
connected to $\exp(-i n \varphi)$, an eigenfunction of the canonical
angular momentum with eigenvalue $-n$.

One can see that $\chi \in {\cal P}_0$ provided $n >0$.
Applying $L$ in the form (\ref{angust}), one sees that the ansatz is
consistent only for $0<e B^0 \theta<1$; we shall study other regimes
later on.  One has,
\be
L \chi = -\left(n + \bar \alpha\right)
\chi  \; , \;\;\;  0<e B^0 \theta<1
\ee
where
\be
\bar \alpha = \frac{\alpha}{1 -eB^0\theta}
\,\,\, .
\ee
One can determine $h( N)$  starting from $n=1$. The corresponding
$\chi$ eigenstate will be denoted as $\chi_{01}$,
\be
\chi_{01} = a^\dagger h( N)
\,\,\, .
\label{eigen}
\ee
Calling
\be
h_m =
\langle m |h( N)|m\rangle \,\,\, ,
\ee
one finds
\be
h_m =\left(\frac{1}{2\pi \theta}
\frac{\Gamma(m + 2 + \bar \alpha )}{(m+1)!\,\Gamma (2 + \bar\alpha)}
(eB^0\theta)^{2 + \bar \alpha}
{\left(1 - e B^0 \theta \right)}{}^m \right)^{1/2}
\,\,\, .
\ee
We can express eigenfunction (\ref{eigen}) in terms of ordinary
functions using the connection $|m\rangle\langle m| \to 2(-1)^m
L_m(2r^2/\theta)\exp(-r^2/\theta)$. We can then explicitly write
\be
\chi_{01}(z,\bar z) = \bar z * \frac{1}{\sqrt\theta}\sum_m
 2(-1)^m h_m L_m(2r^2/\theta)\exp(-r^2/\theta)
 \label{seve}
\,\,\, .
\ee

A tower of states with increasing energy
can be constructed from $\chi_{01}$ by acting with
$c^\dagger$ and $d^\dagger$,
\be
\chi_{kl} = \left(c^{\dagger}\right)^k
\left(d^{\dagger}\right)^{l-1} \chi_{01}  \; , \;\;\; l>k
\,\,\, .
\label{esosno}
\ee
Being $l>k$, one can see that $\chi_{kl}  \in {\cal P}_0$. The
corresponding energy and angular momentum eigenvalues are
\ba
E_{k} &=&  \omega\left(k + \frac{1}{2}\right) \nonumber\\
 && ~ ~ ~ ~ ~ ~ ~ ~ ~ ~ ~ ~ ~ ~ ~ ~ ~ ~ ~ ~ ~ ~ ~ ~0<e B^0 \theta<1 \nonumber\\
L_{kl} &=& (k -l - \bar \alpha)
\,\,\, .
\label{ca}
\ea
It is not difficult to see that in the commutative ($\theta \to
0$) limit, eigenstate (\ref{seve}), and consequently  all the
tower (\ref{esosno}), coincide with the corresponding ``class II''
states found in \cite{JC}.

Inspired by the ordinary space results, one could try to construct
a second tower of (``class I'') states  using, instead of
(\ref{could}), the condition
\be
d \eta = 0
\,\,\, .
\label{dould}
\ee
In contrast with what happens in ordinary space \cite{JC},  in
noncommutative space one finds that there is no solution belonging
to ${\cal P}_0$ in the region $0<e B^0 \theta<1$.

Let us now study the region $e B^0 \theta<0$. In this case $\kappa
<0$ and then $c_1, c_2$ (and also $d_1,d_2$) interchange their
roles of creation and annihilation operators. Now, non-trivial
solutions $\eta$ can be found in this region if one starts from
the condition
\be
d \eta = 0
\,\,\, .
\ee
Eigenstates and eigenvalues can be obtained from a state $\eta_{10}$
of the form
\be
\eta_{10} = a^\dagger f( N)
\ee
with $f( N)$ given by
\be
f_m =\left(\frac{1}{2\pi\theta}
\frac{\Gamma(m + 2 + \bar \alpha )}{(m+1)!\,\Gamma (2 + \bar\alpha)}
\frac{~ ~ (-eB^0\theta)^{2+\alpha}}{\left(1 - e B^0 \theta
\right)^{m-2-\alpha} }
\right)^{1/2}
\,\,\, .
\ee
The tower of solutions with increasing energy is now  given by
\be
\eta_{kl} = \left(c^{\dagger}\right)^{k-1}
\left(d^{\dagger}\right)^{l}  \eta_{10}  \; , \;\;\; k>l
\,\,\, ,
\ee
and eigenvalues take the form
\ba
E_{k} &=&  \omega\left(k + \frac{1}{2} - \bar \alpha\right) \nonumber\\
 && ~ ~  ~ ~ ~ ~ ~ ~ ~ ~ ~ ~ ~ ~ ~ ~ ~ ~ ~ ~ ~ ~ ~ ~ ~ ~
 e B^0 \theta<0  \,\,\, ,\nonumber\\
L_{kl} &=& (l - k - \bar \alpha)
\label{cax}
\,\,\, .
\ea
As in the previous case, eigenstates and eigenvalues coincide, in
the $\theta \to 0$ limit with those called ``class I''  in
ordinary space \cite{JC}.

We have not found an acceptable ansatz for the $eB^0\theta >1$
region. The problem is the following: in order to treat this
region, one has to modify the ansatz for the vector potential,
adding to $A_z$ as given by (\ref{vec}) a term proportional to $z$
(and a term proportional to $\bar z$ in $A_{\bar z}$). Now, with
this form for the vector potential, we were not able to construct
the tower of eigenstates belonging to ${\cal P}_0$. It is
interesting to note that this region does not exist in the
commutative $\theta \to 0$ limit.

The discussion above corresponds to the fundamental representation as
defined by (\ref{deris}). The analysis for the anti-fundamental representation
follows the same steps. One defines operators $c_i$ and $d_i$ as in
(\ref{alpha1})-(\ref{alpha2}) but with the covariant position and derivative
operators in the anti-fundamental, according to (\ref{deris}). Again, we
have to restrict states to a subspace $\tilde{\cal P}_0$
such that, on it, the operator algebra reduces to the
canonical one, eq.(\ref{algebra1}). The condition on states reads now
\be
\tilde\chi \in \tilde {\cal P}_0  \rightarrow
  \tilde\chi| 0 \rangle \langle 0 |= 0 \; , \;\;  \;\;\;
\,\,\, .
\label{formita2}
\ee
One finds for the lowest state \ba \tilde \chi_{01} = s(N) a
\,\,\, . \ea with $s(N)$ adjusted so that $c \tilde \chi_{01} =
0$. From this state, one construct a tower of states through
\ba
\tilde\chi_{kl} &=& \left(c^{\dagger}\right)^k
\left(d^{\dagger}\right)^{l-1} \,{\tilde\chi_{01}}  \; , \;\;\; l>k
\,\,\, ,\nonumber\\
E_{k} &=&  \omega\left(k + \frac{1}{2}\right)  ~ ~ ~ ~
~ ~ ~ ~ ~ ~ ~ ~ ~ ~ ~ ~ ~ ~
0<e B^0 \theta<1 \nonumber\\
L_{kl} &=& - (k -l - \bar \alpha)
\,\,\, .
\label{caa}
\ea
Analogously, from
\ba
\tilde\eta_{10} = t(N) a
\,\,\, ,
\ea
one has
\ba
\tilde\eta_{kl} &=& \left( c^{\dagger}\right)^{ k-1}
\left(d^{\dagger}\right)^l\, {\tilde\eta_{10}}  \; , \;\;\; k>l
 \,\,\, ,\nonumber\\
E_{k} &=&  \omega\left(k + \frac{1}{2} - \bar \alpha\right)
~ ~ ~ ~ ~ ~ ~ ~ ~ ~ ~ ~ ~ ~ ~ ~ ~ ~ ~ ~
 e B^0 \theta<0 \nonumber\\
L_{kl} &=& -(l - k - \bar \alpha)
\,\,\, .\label{cax2}
\ea
Note that these results coincide with those for the fundamental
representation except for the sign in the angular momentum
eigenvalues.

It is interesting at this point to connect our results with those
obtained in ordinary (commutative) space by Johnson and  Canright
\cite{JC} in their discussion of the physics of two anyons in a
uniform magnetic field. Using the analogue of operators $c_i$ and
$d_i$, these authors construct two classes of eigenstates, one for
which the energy does   depend on the vortex flux $\alpha$ (class
I) and the other which exhibits an $\alpha$-independent energy
(class II). Concerning the angular momentum, the two classes have
opposite sign eigenvalues. While in the ordinary Landau problem
all states have energy depending only on the integer Landau index,
when particles have attached a flux tube of strength $\alpha$,
class I states (which, according to their angular momentum
eigenvalue  circulate in a classically ``incorrect'' direction)
have an energy which is shifted by $\alpha$.

How this scenario is modified in noncommutative space? Only one
class of states can be constructed both for the fundamental and
the antifundamental representations. Indeed, in the fundamental,
for  uniform magnetic field in the range $0<eB^0 <1/\theta$, only
class II ($\alpha$-independent energy) states exist while for a
range  $eB^0<0$ the only possible states are class I type with
$\alpha$-dependent energy; concerning the antifundamental
representation, the same phenomenon happens. Now, in the
commutative limit ($\theta \to 0$) the fundamental representation
(with charge $e$) and the anti-fundamental representation (with
charge $-e$) merge,  covering the whole range of  values of $eB^0$
so that we have both classes  of solutions in both regions, thus
recovering Johnson and Canright result.

We have then solved explicitly the spectrum of a Hamiltonian
describing a charged particle in the presence of a background
magnetic field and a vortex in both, the fundamental and the
anti-fundamental representations, for the case $1-eB^0 \theta > 0
$. We have shown that the ``hard-core" condition plays an
important role in the selection of the allowed states, giving only
negative (positive) angular momentum eigenstates for the
fundamental (anti-fundamental) representation. It is well known
that the issue of hard-core condition is related to the problem of
self-adjoint extensions of the Hamiltonian in commutative space.
It would be interesting to explore these ideas in the
noncommutative setting and to study the connection between the
parameter $\theta$ and the self-adjoint parameter, both
dimensionful. Like in the commutative space, one would probably
need to solve in this case the complete second-order Schr\"odinger
differential equation.

As we mentioned in the introduction, in commutative space the
problem of a particle on the background of a vortex is equivalent
to the two-anyon problem. In turn, the problem of ayons can be
analyzed as the quantum mechanics of ordinary particles
interacting with gauge fields whose dynamics is governed by a
Chern-Simons term \cite{F}-\cite{JJJ}. As far as we know, this
line of research has not been yet implemented in noncommutative
space and it would be worthwhile to advance along this line. Also,
it would be interesting to analyze this problem within the
path-integral approach developed in \cite{cs}.

It is not clear if our approach can be extended to the region $1-e
B^0 \theta < 0$, which does not have a commutative limit. We hope
to report on these issues elsewhere.

\vspace{1 cm}

\noindent\underline{Acknowledgements}: This work  partially
supported  by UNLP, CICBA, CONICET (PIP 4330/96), ANPCYT (PICT
03-05179), Argentina.  G.S.L. and E.F.M. are partially supported
by Fundaci\'on Antorchas, Argentina. D.H.C. was partially
supported through an ANPCYT fellowship.


\begin{thebibliography}{99}
\bibitem{jac} R.~Jackiw, {\it Topics in Planar Physics},
Banff NATO ASI {\bf 0191} (1989) 240.
\bibitem{jac2} R.~Jackiw,  Annals Phys. (N.Y.) {\bf 201} (1990) 83.
\bibitem{Sus} L.~Susskind,
hep-th/0101029.
\bibitem{Poly}
A.~P.~Polychronakos,
JHEP {\bf 0104}, 011 (2001)
[hep-th/0103013].
\bibitem{Gubser:2001dz}
S.~S.~Gubser and M.~Rangamani,
JHEP {\bf 0105}, 041 (2001)
[hep-th/0012155].
\bibitem{HR} S.~Hellerman and M. van Raamsdonk, hep-th/0103179.
\bibitem{Lee:2001fk}
B.~Lee, K.~Moon and C.~Rim,
hep-th/0105127.
\bibitem{Morariu:2001qa}
B.~Morariu and A.~P.~Polychronakos,
JHEP {\bf 0107}, 006 (2001)
[hep-th/0106072].
\bibitem{dh}
C.~Duval and P.~A.~Horvathy,
hep-th/0106089.
\bibitem{Hellerman:2001yv}
S.~Hellerman and L.~Susskind,
hep-th/0107200.
\bibitem{Freivogel:2001vc}
B.~Freivogel, L.~Susskind and N.~Toumbas,
hep-th/0108076.
\bibitem{R} A.~El Rhalami, E.~M.~Sahraoui and E.~H.~Saidi,
hep-th/0108096.
\bibitem{np} V.P.~Nair and A.~Polychronakos,  Phys. Lett. {\bf B505}
(2001) 267.
\bibitem{jell} A. Jellal, hep-th/0105303.
\bibitem{GLMR} J.~Gamboa, M.~Loewe, F.~M\'endez and J.C.~Rojas, hep-th/0106125.
\bibitem{BNS} S.~Bellucci, A.~Nersessian and C.~Sochichiu, hep-th/0106138.
\bibitem{luk} J.~Lukierski, P.~Stichel, W.~Zakrzewski, Ann. of Phys. (N.Y.)
{\bf 260} (1997) 294.
\bibitem{hd2} C.~Duval and P.~A.~Horvathy, Phys. Lett. {\bf B479} (2000) 284
\bibitem{dh3} C.~Duval and P.~A.~Horvathy, cond-mat/0101449.
\bibitem{Cha1}
M.~Chaichian, A.~Demichev, P.~Presnajder, M.~M.~Sheikh-Jabbari and A.~Tureanu,
hep-th/0012175.
\bibitem{Cha2:2001pw}
M.~Chaichian, A.~Demichev, P.~Presnajder, M.~M.~Sheikh-Jabbari and A.~Tureanu,
hep-th/0101209.
\bibitem{G3}
J.~Gamboa, M.~Loewe and J.~C.~Rojas,
hep-th/0101081.
\bibitem{JC} M.D.~Johnson and G.S.~Canright, Phys. Rev. {\bf B 41} (1990) 6870.
\bibitem{cor2} D.~Bak, S.K.~Kim, K.-S.~Soh and J.H.~Yee, Phys. Rev. {\bf D64}
(2001) 025018.
\bibitem{J} R.~Jackiw and A.N.~Redlich, Phys. Rev. Lett. {\bf 50} (1983) 555.
\bibitem{Mac} A.H.~MacDonald,  Phys. Rev. {\bf B 30} (1984) 3550.
\bibitem{POL} A.~Polychronakos,  Phys. Lett. {\bf B495}
(2000) 407.
\bibitem{GN} D.J.~Gross and N.~Nekrasov, JHEP {\bf 0007} (2000) 034;
 JHEP {\bf 0010} (2000) 021.
\bibitem{JMW} D.P.~Jaktar, G.~Mandal and S.R.~Wadia, JHEP {\bf 0009}
 (2000) 018.
\bibitem{JKL} J.A.~Harvey, P.~Kraus and F.~Larsen,
JHEP {\bf 0012}  (2000) 024.
\bibitem{F} E.~Fradkin, {\it Field Theories of Condensed Matter systems},
Frontiers in Physics, Addison-Wesley, New York, 1991.
\bibitem{JJJ} See R.Jackiw in {\it Diverse Topics in Theoretical and Mathematical
Physics}  World Sci. Singapore, 1995, pp 465-514 and references therein.
\bibitem{cs} H.~Christiansen and F.A.~Schaposnik, hep-th/0106181
\end{thebibliography}
\end{document}